\documentclass[aps,prb,showpacs,byrevtex,showpacs,amsmath,amssymb,twocolumn]{revtex4-1}
\usepackage{amssymb}
\usepackage{epsfig}
\usepackage{graphicx,subfigure}

\begin{document}

\title{Field-induced Orbital Patterns in Ferromagnetic Layered Ruthenates}
\author{Filomena Forte, Mario Cuoco, and Canio Noce}
\affiliation{CNR-SPIN, I-84084 Fisciano (SA), Italy \\
Dipartimento di Fisica ``E. R. Caianiello'', Universit\`a di Salerno, I-84084 Fisciano (SA), Italy}

\begin{abstract}
We study the evolution of orbital patterns in ferromagnetic
layered ruthenates due to the competition of Coulomb interactions,
compressive $\it{c}$-axis and orthorhombic distortions in the
presence of a polarizing orbital field coupled to the angular
momentum. By means of the exact diagonalization on a 2$\times$2
cluster and a cluster embedded analysis where inter-plaquette
interaction is treated on mean field level, we determine the
ground- state phase diagram. Specifically, we demonstrate that,
via the activation of two or three of $t_{2g}$ local orbital
configurations, an external field applied along different symmetry
directions can lead to inequivalent orbital correlated states.
Starting from an antiferro-orbital pattern, for the easy axis case
an orbital ordered phase is induced, having strong next nearest
neighbors ferro-orbital correlations. Otherwise, a field applied
along the hard axis leads a reduction of local orbital moment  in
a way to suppress the orbital order.

\end{abstract}
\pacs{74.70.Pq, 75.25.Dk, 71.10.-w}
\date{\today} \maketitle


\section{Introduction}

Large Coulomb repulsion, orbital degeneracy and lattice
distortions are widely accepted to be key parameters in
determining the intriguing and fascinating phenomena of transition
metal oxide materials~\cite{Maek04,Tok00,Ima98}. Colossal
magnetoresistance (CMR) in perovskitic manganites is one of the
most significant examples of the subtle competition between
spin-orbital-charge degrees of freedom at the borderline between
different types of
ordered states~\cite{Dag01,Dag03}.\\
Moreover, when electrons in degenerate $d$ shells localize in
Mott-Hubbard or charge transfer insulators, unusual collective
behaviors may emerge as a consequence of the orbital frustration
on the spin degree of freedom. In this context, quantum effects
associated with the orbital degree of freedom are expected to be
of great relevance both for $e_g$~\cite{Fei97,Khal97} and $t_{2g}$
systems~\cite{Kha00,Kha01}.

Recently, the physics of Ca$_3$Ru$_2$O$_7$ compound has attracted
a lot of interest due to a novel type of observed
CMR~\cite{Cao03a,Cao04,Liu99,Lin05,Kar04,Kar06} that is likely to
be driven by orbital rather than quantum spin effects. This
compound belongs to the Ruddlesden-Popper series
Ca$_{n+1}$Ru$_n$O$_{3 n+1}$, $n$ being the number of Ru-O layers
per unit cell, and thus exhibits a bilayered structure. The
crystal structure is orthorhombic with significant distortions in
the $ac$ plane due to the tilting of RuO$_6$ octahedra around the
$b$ axis\cite{Cao00}. In zero field, Ca$_3$Ru$_2$O$_7$ undergoes
an antiferromagnetic (AFM) transition at T$_N$=56~K while
remaining metallic and then a Mott-like transition at
T$_{MI}$=48~K with a dramatic reduction (up to a factor of 20) in
the conductivity for
T$<$T$_{MI}$~\cite{Cao03a,Cao04,Kar04,Ohm04,Lin05}. This
transition is accompanied by an abrupt shortening of the $c$ axis
lattice parameter below T$_{MI}$~\cite{Cao03a,Nelson07}. We notice
that in the AFM state, neutron scattering measurements indicate
that the magnetic moments align ferromagnetically within the
double layer, and antiferromagnetically between the double layers
(A-type AFM structure) along the $c$
axis~\cite{Mcc03,Yoshida05}.\\
Concerning the mechanism behind the CMR in Ca$_3$Ru$_2$O$_7$, the
physics involved is likely to be fundamentally different from that
governing all other magnetoresistive materials. Experiments reveal
that the electronic transport is so deeply interrelated to the
spin-orbital correlations that, if one applies a magnetic field
along the magnetization easy-axis $a$ ~\cite{note}, there occurs a
first order metamagnetic transition to a spin-polarized state,
without leading to a full suppression of the insulating state.
Otherwise, the most favorable magnetoconducting state is driven by
the field directed along the hard-axis $b$, where most likely an
orbital disordered phase regime is obtained, without considerably
affecting the magnetic structure. Indeed, in this regime the
magnetization does not change significantly, thus suggesting that
magnetic scattering is not the primary mechanism responsible for
CMR~\cite{Mcc03,Cao03b,Cao04,Lin05}. These transitions are also
accompanied by anisotropic structural changes and modifications of
the coupling between the electrons and the
lattice~\cite{Lin05,Kar04,Kar06,Nelson07}. Notably, when the field
is applied along the low temperature hard axis, there occurs a
sharp structural change that ties the CMR to an increase of the
$c$-axis lattice parameter ~\cite{Nelson07}. The joining togheter
of CMR, structural changes and metal-insulator transition is a
clear indication of the interplay between spin, charge and lattice
degrees of freedom ~\cite{Qu09}.\\
Earlier theoretical studies on the phenomenology occurring in
Ca$_3$Ru$_2$O$_7$ have been devoted to understand the nature of
the antiferromagnetic metallic state~\cite{Spal95,Duffy97} and the
character of the metal-insulator transition~\cite{Dobr98}. The
crucial role of orbital degrees of freedom and structural
distortions has been recently recognized in the conducting
behavior of Ca$_3$Ru$_2$O$_7$~\cite{Lee07} while it has been
extensively analyzed in the single layered compound Ca$_2$RuO$_4$,
where it has gained a lot of extra interest in the context of
orbital-selective Mott transitions~\cite{Anis02,Neupane09,Medici09,Werner09}.\\
Having in mind the experimental scenario above mentioned for
Ca$_3$Ru$_2$O$_7$, the nature of the ground state within the
layered ferromagnetic insulating state turns out to be a relevant
issue to be addressed. In this case, since the spin and charge
degrees of freedom can be projected out, a suitable microscopic
description has to explicitly include the degeneracy of $t_{2g}$
orbital sector, the compressive $c$-axis and the orthorhombic
distortions of RuO$_6$ octahedra. Then, as for other correlated
insulating states, we base the low energy description on an
effective orbital only model in the regime of strong on site
Coulomb repulsion. Due to the many competing couplings involved in
the problem, the analysis of the orbital correlations within the
different ground states is performed via a systematic approach.
Firstly, the given model Hamiltonian is solved by means of exact
diagonalization (ED) technique on a 2$\times$2 plaquette. This
analysis provides an unbiased insight on the character of the
ground-state configurations as the microscopic interactions are
treated without any approximation and on equal footing. Then,
considering an embedded cluster approach where the intra-plaquette
Hamiltonian is solved via ED while the inter-plaquettes
interaction is treated within a self-consistent Mean-Field (MF)
approximation, we extract the phase
diagram for the allowed broken symmetry configurations.\\
There are two main aspects behind the general complexity of the
problem we faced; i) the Heisenberg-like Hamiltonian for
pseudospin-1 that we considered is marked by directional exchanges
that in two-dimensions, due to the peculiar connectivity via the
oxygen ligands of the $t_{2g}$ centers, lead to antiferromagnetic
correlations that are highly frustrated; ii) the tetragonal and
the orthorhombic crystal field terms act like effective orbital
fields that do not commute between each other. In this respect,
even considering the case of a S=1/2 antiferromagnetic Heisenberg
Hamiltonian, one can notice that the application of two
non-commuting fields would produce a problem that cannot be solved
exactly even in
one-dimension.~\cite{Kurmann81,Kenzelmann02,Dmitriev04} We
underline that our orbital case refers to a spin-1 like system
thus it intrinsically contains extra degrees of freedom whose
interplay with the presence of two-non-commuting fields adds extra
difficulties and different scenarios as compared to the spin-1/2
situation.\\In particular, the octahedral distortions lift the
pristine orbital degeneracy and set orbital patterns that
privilege exchange along the easy axis $a$. This corresponds to
lowering the local number of active degrees of freedom, that is
reduced to the only two orbital configurations that can be
effectively connected along the selected direction. Given this
scenario, we show that the application of the field along the hard
axis restores the local orbital fluctuations among the whole
$t_{2g}$ subspace, by reintroducing the extra configuration that
was quenched by the
balance between orbital exchange and crystal field amplitudes.\\
Special attention is devoted to analyzing the consequences of
activating two or three of the $t_{2g}$ local orbital
configurations on the evolution of orbital order. One of the main
findings is that a polarizing orbital field destabilizes
antiferro-orbital ordered phases, both when applying the field
along the easy or the hard axis. However, we show that depending
on the axial direction, we get two fundamentally different
mechanisms: for a field applied along the easy axis $\it a$, the
effect of the field is mainly to reorient the orbital moments
inducing ferro-orbital correlations in the direction of the
transverse applied field. Due to the orbital configurations
involved, this situation still has the features of an orbitally
ordered state, where nearest-neighbor antiferro-orbital
correlations are strongly suppressed. On the contrary, in the hard
axis $\it b$ case, the activated configuration along $b$ reduces
the amplitude of the ordering orbital moments in zero field up to reach an orbital disordered phase.\\
We finally consider possible connections between the symmetry breaking
associated with the orbital ordered state and the peculiar magneto-conducting behavior of layered ruthenates.\\
The paper is organized in the following way: in the next section
we introduce and write down the microscopic model; in Sec. III we
present, for a $2\times2$ plaquette, the numerical results obtained for the phase diagram as
function of the microscopic parameters stressing the role played
by the purely structural distortions and their interplay with
external magnetic field while in Sec. IV we present embedded mean field calculations
based on exact diagonalization of 2x2 plaquette; the last section contains a discussion of the results
presented and the conclusions.

\section{The effective orbital model}

Based on the assumption that the planar spin configuration is not
substantially modified by the applied field, we consider the
coupling between the field and the local orbital angular momentum
contribution as the dominant one. In particular, starting from
different orbital ordered (OO) ground states at zero field, we
will study how these configurations are modified as a consequence
of the induced orbital ``flip" via the coupling between the
applied field and the orbital angular momentum along different
symmetry in-plane directions. Then, by means of these results, we
will consider the evolution of the ground state under the external
field, and in turn use such investigation to discuss the possible
implications with respect to the anomalous CMR phenomenon observed
in the Ca$_3$Ru$_2$O$_7$.

As mentioned in the Introduction, to investigate the correlations
developing within the orbital sector, we start from a situation
where the system is fully polarized and all the spins are aligned
along a given crystallographic direction, in accordance to the
observed A-type configuration. In this frame, the attention is
focused on the  orbital degree of freedom, out of the competition
between the Coulomb repulsion (orbital exchange), octahedral
distortions, and the effect of the orbital field. In particular,
having in mind the Ca-based ruthenates, the case of flat
octahedral configuration with orthorhombic distortions is
considered. In such a scheme we can explore the interplay between
orbital exchange and the quenching of orbital correlations
provided by structural distortions. The specific model Hamiltonian
we refer to is built up by the following contributions:
\begin{equation}
H=H_{exc}+H_{tet}+ H_{ort}+H_{\alpha}\ , \label{eq:Htotal}
\end{equation}
where $H_{exc}$ stands for the orbital exchange interaction, $H_{tet}$ and $H_{ort}$ are the crystalline fields terms, describing
local $c$-axis tetragonal
distortions and in-plane orthorhombic deformations, respectively, and $H_{\alpha}$ represents the coupling of the orbital angular momentum to the external field.\\
The first term in Eq.~(\ref{eq:Htotal}) is derived from the
superexchange Hamiltonian for $S=1$ spins and it is able to
describe, in the low energy limit, the interaction arising from
the virtual excitation between two neighboring sites. In a
previous work~\cite{Cuo06}, we have proved that such a
description, which has the same roots as the superexchange in a
Mott insulator with nondegenerate orbitals~\cite{Khal97}, can be
applied to the case of ruthenates, being consistent with a
significant value of an on-site intraorbital Coulomb interaction.

\noindent In ruthenates, the Hund's coupling is larger than the
crystal field splitting of $t_{2g}$ subspace and the Ru$^{4+}$
ions are in triplet configuration. Dealing with an atomic
$t_{2g}^4$ configuration, the superexchange interaction arises
from virtual processes $\langle i,j \rangle$ of the type $d_i^4
d_j^4 \rightarrow d_i^5 d_j^3 \rightarrow d_i^4 d_j^4$. Each of
the orbitals involved is orthogonal to one cubic axis; for
instance, $d_{x}$ is orthogonal to the $x$ axis while $d_{y}$ and
$d_{z}$ are orthogonal to $y$ and $z$, respectively. Due to the
connectivity via intermediate oxygen ligands, only two out of the
three $t_{2g}$ orbitals can be effectively connected for each
axial direction.

The spin/orbital model for Ca-ruthenates can be directly derived
from the one adopted for cubic vanadates~\cite{Kha01}, by
performing a particle-hole transformation within the $t_{2g}$
sector, in a way to map the problem of four electrons to that of
two holes and consequently the doublon states in `no hole'
configurations. In this way, we reduce ourselves to consider
virtual excitations of the kind $d_i^2 d_j^2 \rightarrow d_i^3
d_j^1\rightarrow d_i^2 d_j^2$.

\noindent Here, we refer to a simplified version of the
Hamiltonian in Ref.~\cite{Cuo06}, obtained under the assumption $
J_{H}/U \rightarrow 0$, $J_H$ being the Hund's coupling and $U$
the (
large) Coulomb interaction. Moreover, in the present case,
the spin degrees of freedom have been projected out, by assuming a
fully polarized planar configuration, where $\langle S_i \cdot S_j \rangle=1/4$. The orbital exchange
Hamiltonian is then written as follows:
\begin{eqnarray}
H_{exc}=\frac{1}{2} J \sum_{\gamma} \sum_{\langle ij \rangle
\parallel \gamma} [\vec{\tau_i} \cdot
\vec{\tau_j}+\frac{1}{4} n_i n_j]^{(\gamma)}\
,\label{eq:Hexchange}
\end{eqnarray}
where $\vec{\tau}^{\gamma}_i=\{{\tau}^{\gamma}_{x\,i},{\tau}^
{\gamma}_{y\,i},{\tau}^{\gamma}_{z\,i}\}$ are pseudospin
operators, acting in the subspace defined by the two orbitals
which can be linked along a given $\gamma$ direction.\\
For our purposes, it is convenient to characterize the
correlations of each orbital flavor in terms of the `no hole'
states {$\hat x$, $\hat y$, $\hat z$}, standing for the local
orbital configurations having no hole in the corresponding orbital
$d_x$, $d_y$, $d_z$. For instance, if the bond is along the
$z$-axis, the virtual hopping connects $\hat x$ and $\hat y$ only,
and the orbital interaction may be expressed via the Schwinger
representation: $\tau_{+\,i}^c={p^\dagger}_{i y} p_{i x}$,
$\tau_{-\,i}^c={p^\dagger}_{i x} p_{i y}$,
$\tau_{z\,i}^c=\frac{1}{2}(n_{i y}-n_{i x})$, and $n_i^c=2 {n}_{i
z} +{n}_{i y} +{n}_{i x}$, where $n_{i \gamma}={p^\dagger}_{i
\gamma}p_{i \gamma}$ and ${p^\dagger}_{i \gamma}$ and $p_{i
\gamma}$ are creation and annihilation operators for $\hat \gamma$
configuration at site $i$. Similar relations with exchange of bond
index hold for the other axis directions. Note that, to avoid any
confusion with the pseudospin components, we renamed the orbital
flavor $z=c$. Moreover, $J=4 t^2/U$, $t$ being the hopping
amplitude assumed equal for all the orbitals in the $t_{2g}$
manifold, due to the symmetry relations of the connections
via oxygen $\pi$ ligands~\cite{Cuo06,Cuo06bis}.\\
It is worth to notice that, in absence of crystal field terms
lifting orbital degeneracy, orbital exchange described by
Hamiltonian in Eq.~\ref{eq:Hexchange} is intrinsically frustrated:
due to orbital singlets composition, antiferro-orbital
correlations along $x$ preclude exchange along $y$ and viceversa.
The result is a superposition of valence bond states which are
uncorrelated along orthogonal directions.

Concerning the crystalline field contributions, $H_{tet}$ is the
tetragonal term linked to the compressive or tensile character
of the $c$-axis octahedral distortions.
In the `no hole' state representation, $H_{tet}$ is written as follows:\\
\begin{eqnarray}
H_{tet}=\Delta_{tet}\sum_{i} [n_{i z} -\frac{1}{2}(n_{i y} +n_{i
x})]\ . \label{eq:CFtet}
\end{eqnarray}
Eq.~(\ref{eq:CFtet}) implies that negative $\Delta_{tet}$ values
lower the energy of $\hat z$ configurations, thus simulating a compressive distortion.
This case is relevant for describing the short-$c$-axis insulating phase of Ca$_3$Ru$_2$O$_7$\\
The $H_{ort}$ term describes in-plane orthorhombic deformations as
due to a difference between the $a$ and $b$ crystallographic axis.
From a general
point of view, the orthorhombic deformations are introduced by
means of the coupling between the orbital degree of freedom and
the strain fields
$\varepsilon_1=\varepsilon_{xx}-\varepsilon_{yy}$ corresponding to
a rotation axis like $[1 0 0]$ and
$\varepsilon_2=\varepsilon_{xy}$ corresponding to a rotation axis
$[1 1 0]$ in the tetragonal basis $[x y z]$. These distortions
tend to lift the orbital degeneracy and thus can be viewed as
effective orbital polarizing fields. From symmetry
considerations\cite{Anis02,Sig04}, one can show that a general
form for the orthorhombic term can be expressed as follows:
\begin{eqnarray}
H_{ort}=  \sum_{i} (\Delta_{1,o} \tau^c_{z\,i}+\Delta_{2,o}
\tau^c_{x\,i})  \,.
\end{eqnarray}
where $\Delta_{1,o}$ and $\Delta_{2,o}$ correspond with the strain
fields $\varepsilon_1$ and $\varepsilon_2$. Having in mind the
orthorhombic distortions for the Ca$_3$Ru$_2$O$_7$ compound, where
the RuO$_6$ octahedra are severely tilted in the $ac$ plane thus
rotated with respect to the $b$ axis, one can assume that the
$\Delta_{2,o}$ term is the relevant contribution to be included.
Therefore, in our analysis the last term in Eq.(1) is just the
following:
\begin{eqnarray}
H_{ort}= \Delta_{o} \sum_{i} \tau_{xi}^{c}\ .
\end{eqnarray}

Still, negative values of $\Delta_{o}$ in $H_{ort}$, tend to
stabilize a quantum state given by a superposition $\sim \hat
x+\hat y$. This corresponds to select a preferential axial
direction for orbital connectivity in the ($a, b, c$) frame that
is obtained by performing a $(\pi/4,\pi/4,0)$ rotation of
$(x,y,z)$ in real space. In the following we will refer to
this as the `orthorhombic' frame and to the corresponding orbitals
as $d_a$, $d_b$, $d_c$.\\
Being interested in the response to an
orbital only coupled field, we also include in the Hamiltonian of
Eq.~(\ref{eq:Htotal}) the coupling between the external field and
the local angular moment\cite{Yos}, i.e.
\begin{eqnarray}
H_{\alpha}= 2 B \sum_{i} \tau_{yi}^{\alpha}\ ,
\end{eqnarray}
where $B$ is the magnetic field coupled to the orbital degrees of
freedom only in unit of Bohr magneton and $\alpha=a, b$ defines
the direction, along which $B$ is applied.

\section{Ground-state orbital patterns for a $2\times2$
plaquette}

In this section we present the ground-state (GS) diagrams,
obtained by means of an exact diagonalization study of the
Hamiltonian above introduced on a 2$\times$2 plaquette. As a
strategy, we follow the evolution of the GS induced by the $c$
axis compressive distortions, orthorhombic deformations and the
external orbital only field. The orthorhombic field is assumed in
a way that the $a$ direction is the easy-axis for the angular
moments.

\begin{figure}[!t]
    \begin{center}
    \includegraphics*[width=0.9\columnwidth]{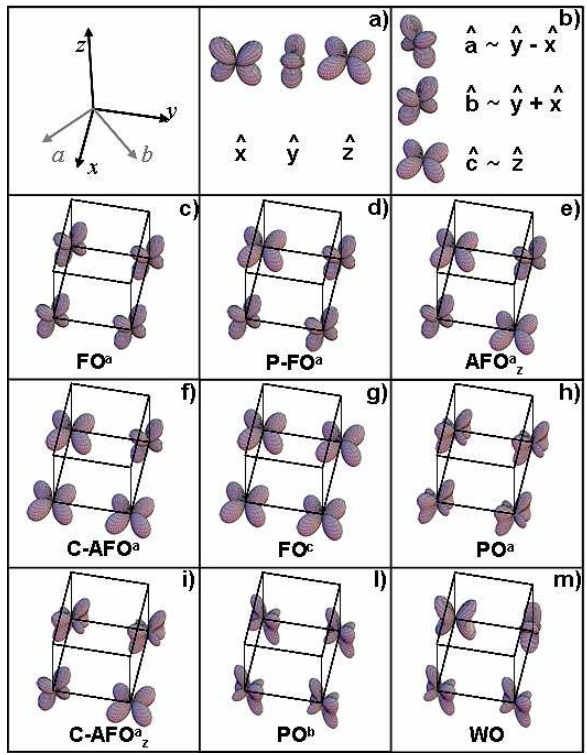}
    \end{center}
\caption{(Color online) Schematic representation of the possible
orbital patterns on a 2$\times$2 plaquette, generated by the
interplay between orbital exchange, tetragonal and orthorhombic
distortions, applied orbital field. First panel shows the axis
notation, panel a) and panel b) the graphical representation of
$\hat x, \hat y,\hat z$ and $\hat a, \hat b,\hat c$
configurations, respectively. Panels c)-m) report the planar
orbital pattern in terms of the local distribution of `no-hole'
configurations; FO$^{\alpha}$/AFO$^{\alpha}$ stand for orbital
patterns having homologue configurations on neighbors sites, for
$\alpha=a, b, c$ flavor. P-FO stands for configurations with
partial FO character, C-AFO for canted AFO. PO$^\alpha$ indicates
`para'-orbital states where where all pseudospins are
$\tau_y^\alpha$ oriented; WO stands for configurations
characterized by a softening of AFO$^a$ correlations, due to
nonvanishing $\langle \tau_y^b \rangle$.}
    \label{fig:configurations}
\end{figure}

It turns out to be easier to characterize the different
ground-state behaviors by evaluating the density and orbital
correlators in the orthorhombic frame. We therefore introduce `no
hole' configurations {$\hat a$, $\hat b$, $\hat c$}, where the
empty state is localized at $d_a$, $d_b$, $d_c$, respectively. In
Fig.~\ref{fig:configurations} b), graphical representation of
{$\hat a$, $\hat b$, $\hat c$}, together with original {$\hat x$,
$\hat y$, $\hat z$} (Fig.~\ref{fig:configurations} a) is reported.
In terms of this new notation, it is easy to define the action of
the pseudospin operators in the orthorhombic basis for the $a, b,
c$ flavors. For example, along the $c$-axis, the pseudospin in the
orthorhombic frame acts on ${\hat{b}}$ and ${\hat{a}}$ and it can
be expressed via $\tau_{+\,i}^c={p^\dagger}_{i b} p_{i a}$,
$\tau_{-\,i}^c={p^\dagger}_{i a} p_{i b}$,
$\tau_{z\,i}^c=\frac{1}{2}(n_{i b}-n_{i a})$.

Finally, in order to simplify the reading of the phase diagrams,
we introduce the following notation: FO, AFO, WO stand for
ferro-orbital (`no-hole' state on neighbors homologue orbitals),
antiferro-orbital (`no-hole' state on neighbors off-diagonal
orbitals) and weak antiferro-orbital correlations, respectively. Besides, P-FO stands for
configurations with partial FO character, C-AFO for canted AFO while PO indicates the para-orbital high-field configuration. Furthermore, the superscripts ($a$, $b$, $c$) stand for the
direction of the main contributions for the orbital correlators in
terms of the pseudospins operators. The underscript $(x,y,z)$
refers to the main character of the orbital correlations in the
pseudospin space, i.e. it provides indication of the anisotropy of
the orbital pattern. A schematic view of the representative
orbital pattern configurations that contribute to the ground state
on the 2$\times$2 plaquette is reported in
Fig.~\ref{fig:configurations} c)-m), where the main component of
the ground state is represented by means of the local density
distribution of `no-hole' configurations, in a way that a specific
orbital on a given site stands for the local configuration having
no holes in that orbital.\\
We notice that, except for the FO/AFO regions that are expected
without explicit calculations, all the remaining phases are
untrivial since they show an intermediate character, where the
correlations have competing behaviors. For completeness, here we
specify the way those regions are classified: i) PO stands for the
para-orbital uncorrelated high-field-configuration, driven by the
field H$_\alpha$, where the pseudospins are $\tau^\alpha$ oriented
and the correlators $\langle \tau^{\alpha}_i \tau^{\alpha}_j
\rangle$, evaluated in the orthorhombic frame, have the maximum
saturated value of 0.25 [see for example Fig. 10 a) and 12 b)];
ii) generic C-AFO stand for regions where nearest-neighbor
antiferro-orbital correlations characterizing AFO are strongly
suppressed and coexist with FO components, as shown for example in
Fig 10. iii) WO has the feature of an isotropic weak
antiferro-orbital phase, obtained when the field is applied along
the hard axis $b$, resulting from a reduction of the
nearest-neighbors $\langle \tau^a_{yi} \tau^a_{yj} \rangle$
antiferro-type correlations due to a non-vanishing onsite momentum
$\tau^b_y$. In this region, FO and AFO correlations are the same
order of magnitude, as shown in Fig. 12, meaning that no preferred
behavior can be identified.

\begin{figure}[!h]
    \begin{center}
     \includegraphics*[width=1\columnwidth]{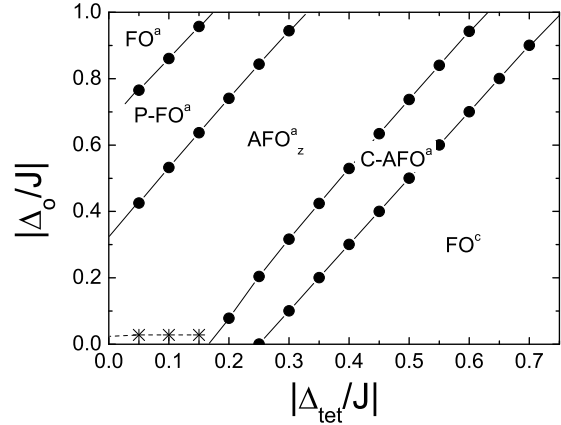}
    \end{center}
\caption{Diagram summarizing the behavior of the GS configurations
of a $2\times2$ cluster as a function of $\Delta_{o}/J$ and
$\Delta_{tet}/J$. Stars stand for smooth crossovers in the
relevant correlators, solid dots for abrupt jumps. Phases are
labelled in accordance to Fig.~\ref{fig:configurations}.}
    \label{fig:PDDtetDort}
\end{figure}

\subsection{Effect of structural distortions}

As a first step we discuss the case at zero applied field. In
Fig.~\ref{fig:PDDtetDort}, the ground-state diagram is shown,
obtained by varying the crystal field microscopic parameters
$\Delta_{tet}$ and $\Delta_{o}$ in unit of $J$. It turns out to be
convenient to span the different regions of
Fig.~\ref{fig:PDDtetDort} by fixing an ideal horizontal line
corresponding to e.g. $|\Delta_{o}/J|=0.8$, while varying the
tetragonal parameter.\\
Tetragonal crystal field acts as an effective orbital bias that
tunes the density of $\hat{c}$ configurations from 0 to 1,
corresponding to weak and strong regime of compressed octahedra,
respectively. Looking at the relevant correlation functions, we
see that the two ending regions are both characterized by a
complete quenching of the orbital correlations. In the weak
tetragonal regime, the orthorhombic field is strong enough to give
rise to a complete FO$^{a}$ (see Fig.~\ref{fig:configurations},
panel c)), where stabilizing ${\hat{b}}$ configurations generates
full FO correlations for the ${\tau^{a}_{z}}$ components. In the
extreme flattened RuO$_6$ regime, instead, we get a FO$^{c}$
region (see Fig.~\ref{fig:configurations}, panel c)), where the
ferro-orbital correlations are saturated along the $z$ direction
for the flavor $c$.\\
These two regions are linked by intermediate regions characterized
by a partial filling of the the $t_{2g}$ subspace. We highlight
that, due to the constraint of the planar geometry, and that of
the orthorhombic field favoring $\hat{b}$ configurations, the
orbital interaction along $b$, involving $\hat{a}$ and $\hat{c}$,
is inactive thus yielding an anisotropic
effect between $a$ and $b$ directions.\\
Particularly, in the regime of intermediate tetragonal field,
0.2$<|\Delta_{tet}/J|<0.4$, we find an AFO$^{a}_{z}$ region,
represented in Fig.~\ref{fig:configurations} e), characterized by
the alternation of $\hat{c}$ and ${\hat{b}}$, that gives rise to
predominant AFO correlations along $z$ direction, for flavor $a$.
This region is interesting when compared with the magnetic/orbital
pattern in Ca$_3$Ru$_2$O$_7$; indeed, the occurrence of an AFO
pattern can explain, in a spin/orbital picture, the existence of
ferromagnetic in-plane correlations, and it can be also related to
the poorly metallic behavior of the GS, as due to an OO pattern.\\

We finally remark that, in the AFO$^{a}_{z}$, the order of
magnitude of the $x, y, z$ correlators for the $a$ flavor is
comparable, though the amplitude for the $z$ channel is larger
than the others.

\subsection{Interplay between structural distortions and orbital
polarizing field}
\subsubsection{Easy axis}

Let us now discuss the interplay between the structural
distortions and the effects of a polarizing field. In doing that, we consider
just one representative value for the orthorhombic strain
$|\Delta_{o}/J|=0.4$. This is the most interesting regime where the structure of the ground state is more sensitive to external small perturbations. This would permit to explore the
modifications induced on the corresponding line of
Fig.~\ref{fig:PDDtetDort}, and particularly of the region at the
boundary between AFO$^a$ and FO$^{c}$, occurring for
intermediate/strong tetragonal crystal
field amplitude.\\
\begin{figure}[!t]
    \begin{center}
   \includegraphics*[width=1\columnwidth]{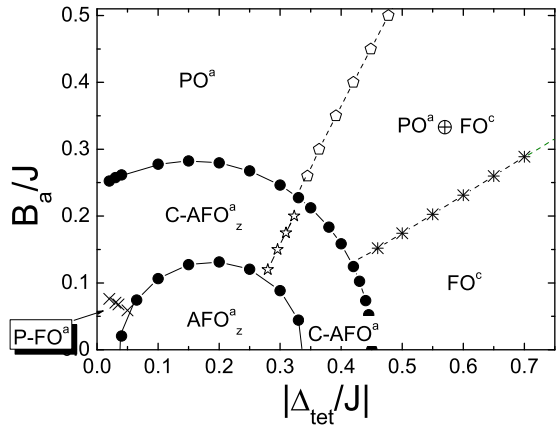}
    \end{center}
\caption{Ground-state diagram of a $2\times2$ cluster obtained for
an orbital-only applied field along the easy axis $a$ as a
function of $|\Delta_{tet}/J|$, for a fixed value
$|\Delta_{o}/J|$=0.4. Dots correspond to abrupt jumps in the
evolution of the relevant off-site orbital correlators; stars and
diamonds to smooth crossover in the correlation functions.
Different regions are labelled according to
Fig.~\ref{fig:configurations}.}
    \label{fig:PDHeasy}
\end{figure}
Due to the anisotropy introduced by the orthorhombic field, we
distinguish between the response along the $a$ and $b$ directions.
The corresponding ground-state diagrams are displayed in
Figs.~\ref{fig:PDHeasy} and \ref{fig:PDHhard}. The investigation
of these diagrams will be confined to the low fields regime since,
approaching high fields, the orbital momentum becomes fully
polarized, as in the $PO^a$ region of
Fig.~\ref{fig:configurations} h), resulting from stabilizing $\sim
(\hat{b}+i\hat{c})$, or the $PO^b$ region of
Fig.~\ref{fig:configurations} l), corresponding to $\sim (\hat{c}-i
\hat{a})$ local configurations.  We do
believe that, in this limit, the assumption of neglecting the spin magnetization
becomes weak and thus an analysis of the complete
spin-orbital effective model is required.

Fig.~\ref{fig:PDHeasy} shows the occurrence of crossovers in the
orbital correlators among different GS configurations generated
for $B||a$. We do notice a peculiar evolution of the boundary for
the AFO$^{a}_{z}$ state. Indeed, aligning the moment along $y$ for
the $a$ flavor, regions AFO$^a$ and FO$^{c}$ move towards a new
regime where AFO$^a$ along $y$ is partially reduced evolving into
a canted like state C-AFO$^{a}_z$ (see
Fig.~\ref{fig:configurations}, panel i)), with a non-zero total
angular momentum, directed along $y$, coexisting with partial
AFO$^a$ correlations along $z$.

\subsubsection{Hard axis}

\begin{figure}[!t]
    \begin{center}
    \includegraphics*[width=1\columnwidth]{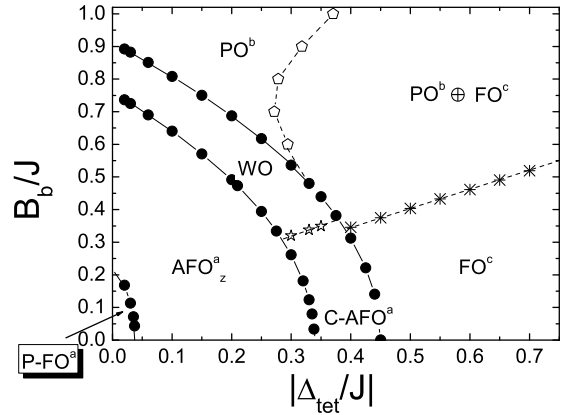}
    \end{center}
\caption{Ground-state diagram of a $2\times2$ cluster obtained for
an orbital-only applied field along the hard axis $b$ as a
function of $|\Delta_{tet}/J|$, for a fixed value
$|\Delta_{o}/J|$=0.4. Dots correspond to abrupt jumps in the
evolution of the relevant off-site correlators; stars and diamonds
to smooth crossover. The regions are labelled according to
Fig.~\ref{fig:configurations}.}
    \label{fig:PDHhard}
\end{figure}
\begin{figure}[!b]
    \begin{center}
    \includegraphics*[width=0.9\columnwidth]{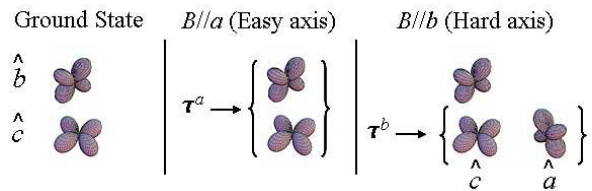}
    \end{center}
\caption{(Color online) Schematic representation of orbital
configurations activated by an applied orbital field. Left panel
shows ${\hat{b}}$ and ${\hat{c}}$ orbital configurations set by
the interplay between crystal field amplitudes. Middle panel shows
the orbital subspace where $\tau^a$ is active; right panel shows
the orbital subspace where $\tau^b$ is active. The orbital field,
applied along hard axis $\textit{b}$, restores orbital
fluctuations among the whole orbital subspace by re-introducing
$\hat{a}$ configurations.}
    \label{fig:TauAction}
\end{figure}

The response for $B||b$ is instead more stiff (see
Fig.~\ref{fig:PDHhard}). In Fig.~\ref{fig:TauAction}, we give a
schematic representation of the action of pseudospin operators on
different orbital subspaces: the ground state at zero field is a
combination of $\hat b$ and $\hat c$ states and the application of
the field along $b$ induces $\hat a$-type configurations that are
unfavored with respect to the orthorhombic deformations. This
implies that a much higher value of the field is required to
activate the orbital exchange along $b$. Along this symmetry
direction, the orbital polarizing field tends to restore a
description where all the three t$_{2g}$ flavors are active. Due
to the new configurations ${\hat{a}}$, and to the competing
mechanisms involved, the resulting effect is a weakly ordered WO
configuration. Such state is marked by a coexistence between AFO
correlations along $a$ and FO along $b$ (see
Fig.~\ref{fig:configurations}, panel m)).
Now, comparing the response of the system for the different field
directions in the interval of crystal field given by about
0.1$<|\Delta_{tet}/J|<0.5$, one can observe that, if the field is
along the $a$ axis, the closest configurations that can be
activated, starting from AFO$^a$ and FO$^{c}$, still have an
orbital ordered pattern (OO) as given by C/AFO$^a$. On the
contrary, for $B||b$, due to the activation of the $b$ flavor, we
get a changeover from an OO pattern to an WO configuration. \\
This analysis  shows that when AFO and FO configurations are close in energy, the
application of the external ?field may act as a tunable parameter
within the orbital configurations, thus suggesting that the complex
field response in the bilayered ruthenate may be consequence of the
subtle interplay between different microscopic mechanisms that involve
the orbital couplings and the fields associated with the octahedral
deformations in presence of large Coulomb coupling. One of the main
findings is the occurrence of an orbital ordered pattern at the
boundary between FO and AFO states, stabilized when the field is
along the easy axis, as due to the effective freezing of one
orbital flavor. When the field is applied along the hard axis,
quantum fluctuations associated with the three-orbital degrees of
freedom are restored and tend to destroy the orbital ordering.

\section{Cluster embedded calculation}
\begin{figure}[!t]
    \begin{center}
    \includegraphics*[width=0.8\columnwidth]{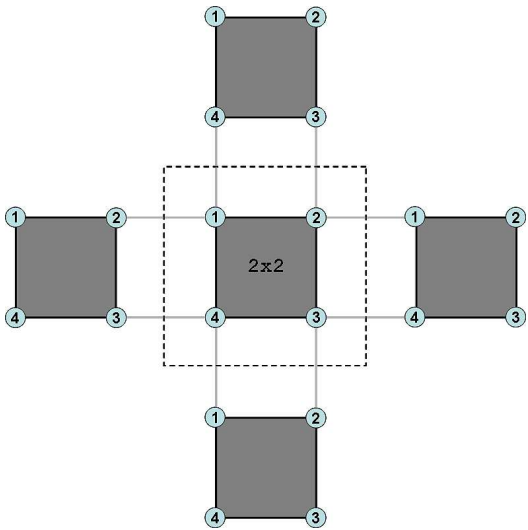}
    \end{center}
\caption{Schematic representation of a 2$\times$2 cluster embedded
in a periodic environment. Filled square denotes 2$\times$2
plaquette described by Eq.~\ref{eq:Htotal}, whose interaction with
the field produced by four surrounding empty squares is expressed
by Eq.~\ref{eq:Hembedded}. Dashed square defines the basic unity
of the periodic structure.}
    \label{fig:clusterembdedded}
\end{figure}

In this section we extend the cluster calculation considering the
case of a 4-site plaquette embedded in a periodic environment as
schematically represented in Fig.~\ref{fig:clusterembdedded}.
Here, the intra-plaquette Hamiltonian is solved via exact
diagonalization and the inter-plaquettes interaction is treated
within a self-consistent Mean-Field approximation. \\
Indeed, the MF scheme of analysis corresponds to decouple the
inter-plaquette terms of the Hamiltonian as it follows:
\begin{eqnarray}
H_{exc}=J \sum_{\gamma=x,y} \sum_{\mu=x,y,z} \sum_{\langle ij
\rangle
\parallel \gamma} [\tau^{\mu}_i \langle \tau^{\mu}_j \rangle + \frac{1}{4} n_i \langle n_j
\rangle]^{(\gamma)}- \nonumber \\
 J \sum_{\gamma=x,y} \sum_{\mu=x,y,z}
\sum_{\langle ij \rangle
\parallel \gamma} [\langle \tau^{\mu}_i \rangle \langle \tau^{\mu}_j
\rangle+ \frac{1}{4} \langle n_i \rangle \langle n_j
\rangle]^{(\gamma)}\, \label{eq:Hembedded}
\end{eqnarray}
where $\gamma$ stands for the axial direction and $\mu$ for the
pseudospin component. Hence, Eq. (\ref{eq:Hembedded}) represents
the exchange interaction between the plaquette and the nearest
neighbors sites in the $(x,y)$ plane, treated on a mean field
level. Here, the average $\langle ...\rangle$ of the local
pseudospin operators is determined after solving the full quantum
problem within the plaquette. The decoupling of the orbital
exchange on nearest-neighbour plaquettes tends to prefer broken
symmetry states with AFO correlations. This tendency is in
competition with the local distortive fields due to the octahedral
distortions and is strongly related to the intra-plaquette quantum
effects. Hence, the solution is determined in a self-consistent
fashion, according to the following scheme: we assign initial
conditions for $\langle \vec{\tau}_i^\gamma \rangle$ and $\langle
n_i^\gamma \rangle$, and use them to evaluate improved expectation
values; the procedure runs until convergence is achieved, with the
requested accuracy. This scheme is applied for all the different
initial conditions associated with possible orbital states
emerging from ED on the 4-site plaquette. In particular, we
consider all the starting configurations where local pseudospins
are oriented along a given direction and FO, A-FO, C-AFO and WO
correlated, and we finally choose the lowest energy orbital
configuration among the ones self-consistently obtained. Mean
field drives the solution towards broken symmetry states having a
specific orbital pattern that can be characterized by local order
parameters $\langle \vec{\tau}_i^\gamma \rangle$, and by
explicitly evaluating orbital correlations developing along
different axial direction of the orthorhombic frame. We point out
that this feature is not intrinsic in the MF approach but is
strongly related to the anisotropic microscopic environment due to
crystal field terms. The latter break rotational invariance in the
pseudospin space, allowing solutions with a specific orbital
patterns to be stabilized.\\ We finally remark that, by exploring
broken symmetry states with  defined OO patterns, we can get
informations about the thermodynamic stability of the
corresponding OO phases.
\subsection{Effect of structural distortions}
\begin{figure}[!t]
    \begin{center}
    \includegraphics*[width=1\columnwidth]{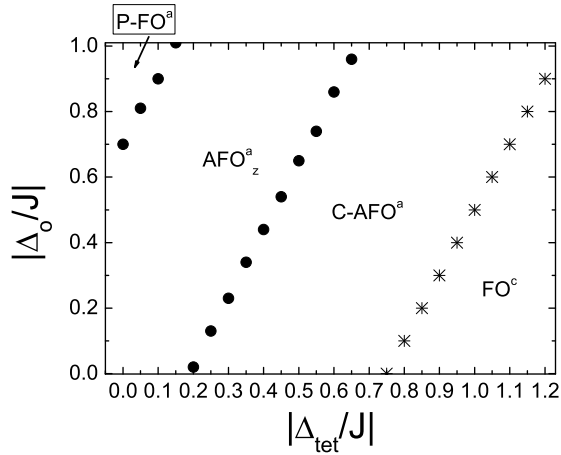}
    \end{center}
\caption{Diagram summarizing the behavior of the GS configurations
as a function of $\Delta_{o}/J$ and $\Delta_{tet}/J$ for a cluster
embedded in a periodic environment.}
    \label{fig:PDDtetDortCE}
\end{figure}

We first consider the MF evolution starting from orbital
configurations of a $2 \times 2$ plaquette at zero applied field.
We look for most stable solutions, chosen among those having
FO-AFO and C/AFO character. In Fig.~\ref{fig:PDDtetDortCE}, we
show the phase diagram as a function of the tetragonal and
orthorhombic crystal field amplitudes. When comparing this diagram
to the cluster calculation (Fig.\ref{fig:PDDtetDort}), the
following features emerge:\\
i) the value of $|\Delta_{tet}/J|$ separating AFO$^{a}_z$ and
C-AFO$^a$ regions is not modified by the MF and scales with
$|\Delta_{o}/J|$ according to the same linear law;\\
ii) in the moderately/highly flattened region, the most stable
solution has a C-AFO$^a$ character. Looking at orbital
correlations in the orthorhombic frame, this corresponds to AFO
correlations in $x,y$ components of flavor $a$ coexisting with FO
correlations for $\tau_z^a$. Approaching the extremely flattened
side, the AFO correlations for $\tau_{x,y}^a$ are gradually
replaced by the FO character of $\tau_z^a$ $\tau_z^b$, which is
saturated reaching FO$^c$.
Moreover $\tau_x^c$ is decreasing and FO-correlated.\\
iii) In the regime of strong orthorhombic field, we observe that
the full FO$^a$ is no more stable and leaves the place to a region
having a uniform P-FO$^a$ character (the evolution of the
expectation values as a function of the orthorhombic field suggests
that FO$^a$ is stabilized for higher values of $|\Delta_{o}|
\gtrsim 2$). This configuration is somehow symmetric with respect
to C-AFO$^a$, but has an inverted balance between $|\hat z
\rangle$ and $(\hat x,\hat y)$ sector.\\
iv) in the AFO$^a_z$ the self-consistence leads to a state with
pure $z$ components of the $a$ and $b$ flavor (correlations along
$x$ and $y$ renormalize down to zero while they were not vanishing
for the uncoupled cluster case.)
\subsection{Effect of orbital-only applied field}
\begin{figure}[!t]
    \begin{center}
    \includegraphics*[width=0.9\columnwidth]{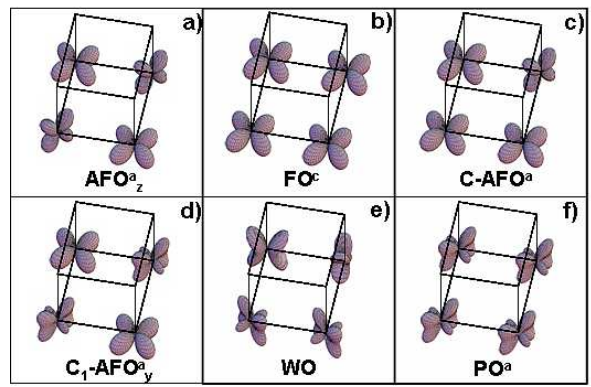}
    \end{center}
\caption{(Color online) Schematic representation of possible
orbital patterns for a $2\times2$ cluster embedded in a periodic
environment as due to the interplay between orbital exchange,
tetragonal and orthorhombic distortions, and applied orbital
field. Panels a)-f) report the planar orbital pattern in terms of
the local distribution of `no-hole' configurations;
FO$^{\alpha}$/AFO$^{\alpha}$ stand for orbital patterns having
homologue configurations on neighbors sites, for $\alpha=a, b, c$
flavor. C-AFO for canted AFO. PO$^\alpha$ indicates `para'-orbital
states where where all pseudospins are $\tau_y^\alpha$ oriented;
WO stands for configurations characterized by a softening of
AFO$^a$ correlations, due to nonvanishing $\langle \tau_y^b
\rangle$ }
    \label{fig:configCE}
\end{figure}
\begin{figure}[!b]
    \begin{center}
    \includegraphics*[width=1\columnwidth]{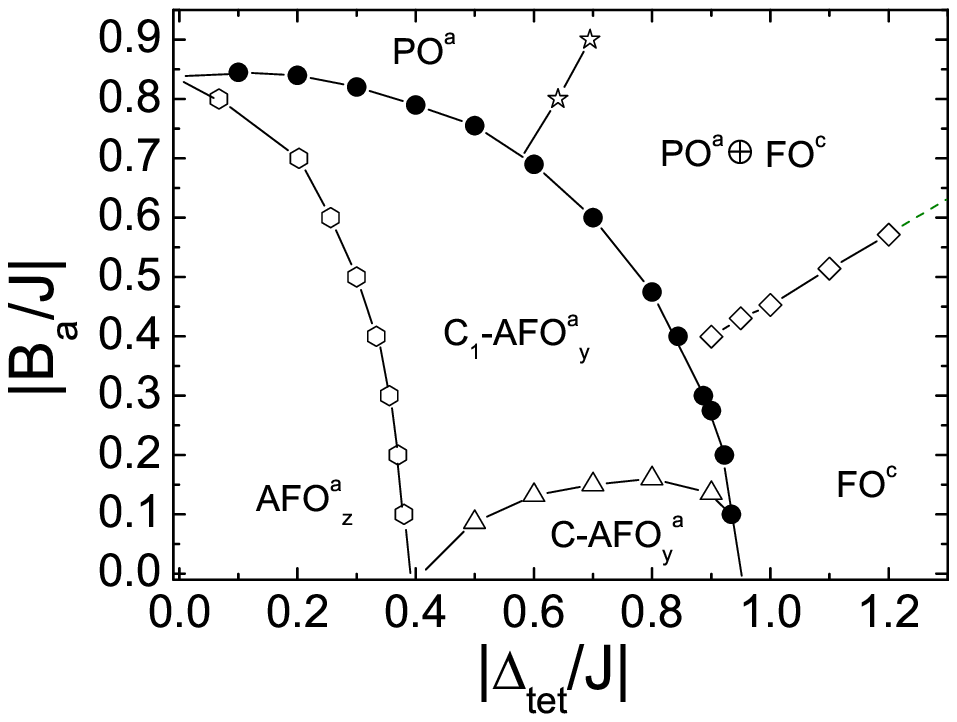}
    \end{center}
\caption{Diagram summarizing the behavior of the GS configurations
as a function of $\Delta_{tet}/J$ and $B_a/J$, for a cluster
embedded in a periodic environment at $|\Delta_o/J|=0.4$.}
    \label{fig:PDHeasyCE}
\end{figure}
\subsubsection{Easy axis}
\begin{figure}[!t]
\begin{center}
\begin{tabular}{c}
\begin{minipage}{\columnwidth}
\includegraphics*[width=1\columnwidth]{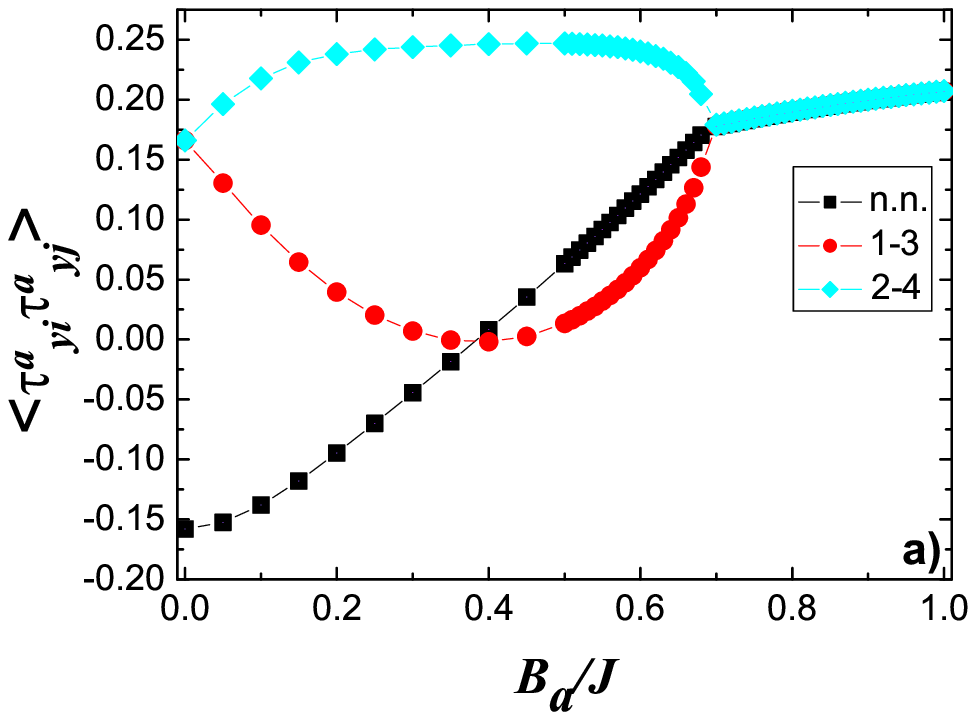}
\end{minipage}
\\
\begin{minipage}{\columnwidth}
\includegraphics[width=1\columnwidth]{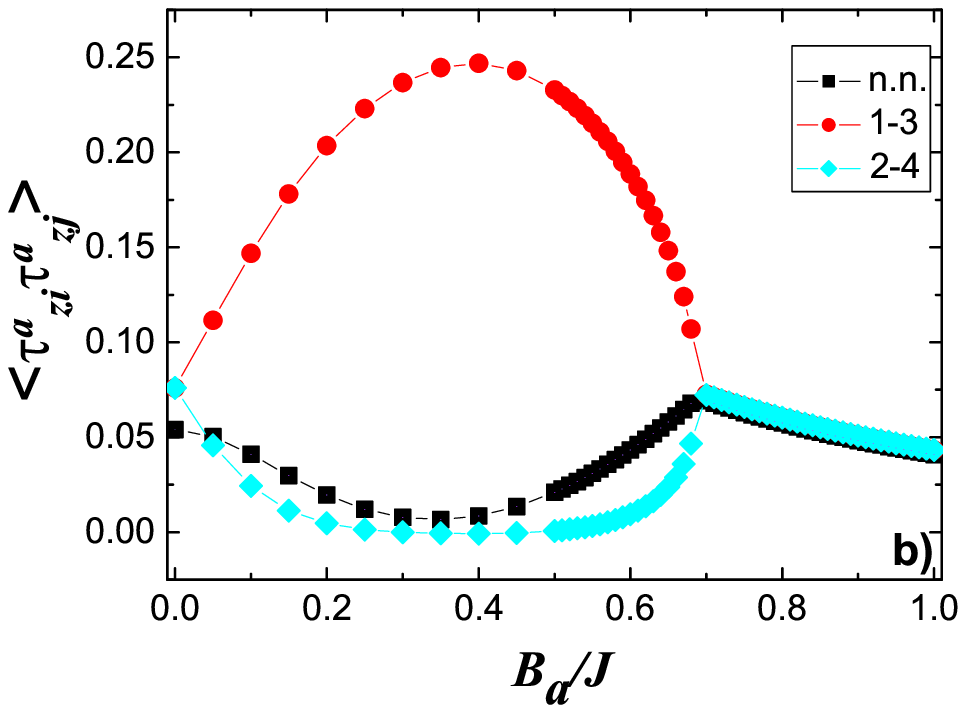}
\end{minipage}
\end{tabular}
\end{center}
\caption{(Color online) Expectation value of orbital correlators
in the orthorhombic frame as a function of applied field, evaluated
for $|\Delta_{tet}/J|=0.6$ and $|\Delta_{o}/J|=0.4$. Panels a)-b)
show dominant contributions arising when the field is applied
along easy axis.} \label{fig:correlatorsEasyCE}
\end{figure}
The effect of a polarizing field along easy axis $a$ is to induce
a nonvanishing $y$ component of local orbital momentum $\langle
\tau_y^a \rangle$. This anisotropic effect is thus relevant for
the evolution of the C-AFO$^a$, where AFO$^a$ correlations exist
both for the $x$ and $y$ components. Therefore, we expect that the
polarizing field naturally wil select the latter. Moreover it's
interesting to see how the induced FO correlations compete with
the general tendency towards AFO. In Fig.~\ref{fig:PDHeasyCE}, we
report the evolution of the line $|\Delta_{o}/J|=0.4$ of
Fig.~\ref{fig:PDDtetDortCE}, as a function of the applied field.
As one can see, one of the main differences, with respect to the
4-site plaquette calculation (see Fig.~\ref{fig:PDHeasy}), is that
now most of the transitions occur as a smooth crossover in the
relevant correlators, represented by empty dots and stars (see
Fig.~\ref{fig:correlatorsEasyCE}).\\
High field regions are
uninteresting (PO phases are stabilized), so we skip them in the
discussion. In the low $|\Delta_{tet}/J|$ region, we see that the
AFO$^a_z$ region remains quite stable and the effect of the
polarizing field is to progressively orient the local orbital
momentum from $z$ to $y$. The main modifications occur in the
moderately flattened region where the C-AFO region--that now has
broken symmetry, being active only for $y$ component--leaves place
to C$_1$-AFO$^a$$_y$. This region has the features of a
strengthened OO, where $\hat c$ and $\hat a$ configurations
alternate along the bonds [see Fig. \ref{fig:configCE} d)].
Looking at orbital correlators in Fig.~\ref{fig:correlatorsEasyCE}
a) and b), we deduce that in this region the competition between
crystal field amplitude and applied field has the effect to
suppress nearest neighbors exchange. The OO that is stabilized has
a strong FO character along opposite diagonals for $\tau^a_y$ and
$\tau^a_z$, respectively, as shown by next nearest neighbors
correlations, which are almost saturated to a value of $\sim
0.25$.\\
To clarify the mechanism leading to the suppression of AFO
correlations, we point out that the characteristic energy scale is
of the order of $J$. The analogy with an Heisenberg
antiferromagnet, disordered by an applied magnetic field, suggests
that orbital correlations are tuned by the interplay between
applied field and orbital exchange, in a way that is not dependent
on the crystal field amplitude, once the orthorhombic deformations
have selected the `easy flavor' $a$.
\begin{figure}[!t]
    \begin{center}
    \includegraphics*[width=1\columnwidth]{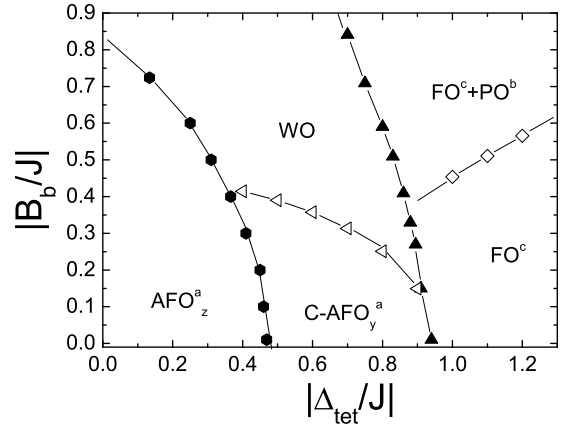}
    \end{center}
\caption{Diagram summarizing the behavior of the GS configurations
as a function of $\Delta_{tet}/J$ and $B_b/J$, for a cluster
embedded in a periodic environment at $|\Delta_o/J|=0.4$.}
    \label{fig:PDHhardCE}
\end{figure}
\begin{figure}[!t]
\begin{center}
\begin{tabular}{c}
\begin{minipage}{\columnwidth}
\includegraphics*[width=1\columnwidth]{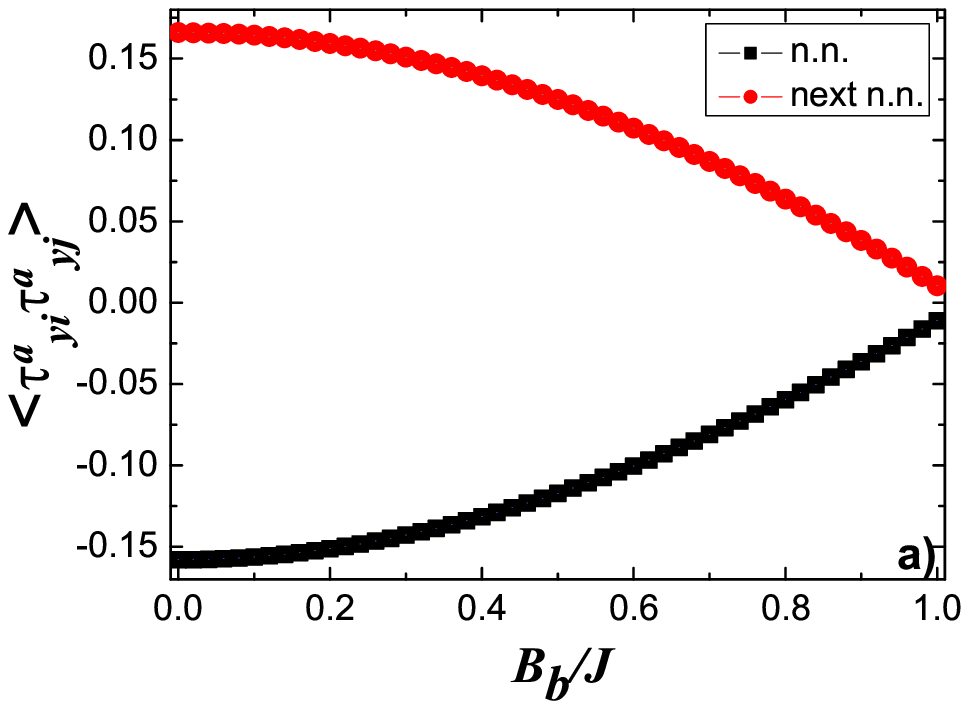}
\end{minipage}
\\
\begin{minipage}{\columnwidth}
\includegraphics[width=1\columnwidth]{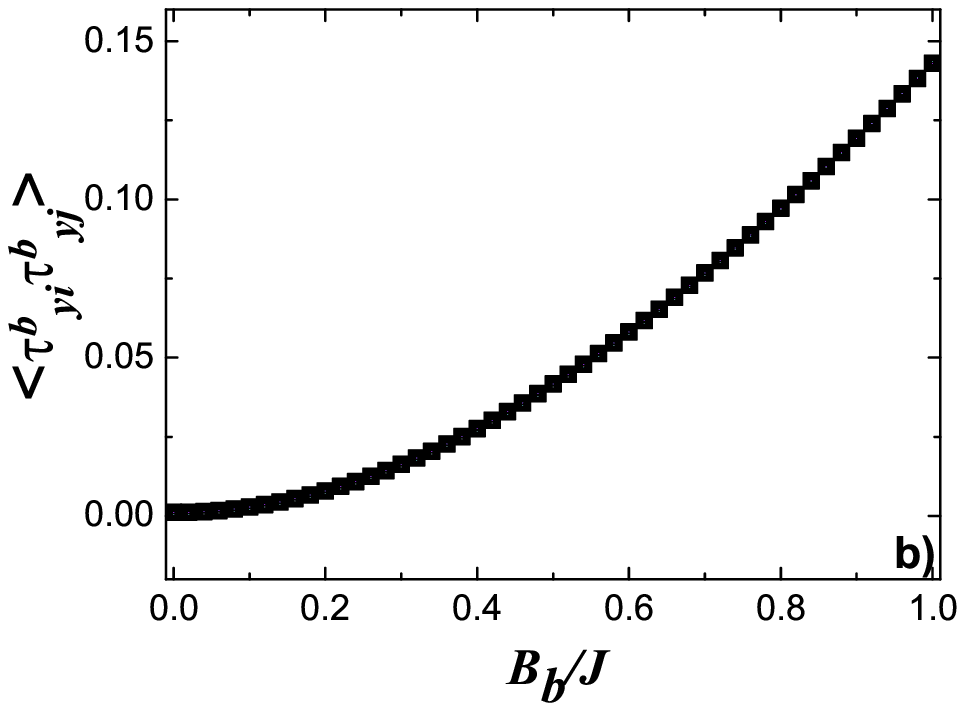}
\end{minipage}
\end{tabular}
\end{center}
\caption{(Color online) Expectation value of orbital correlators
in the orthorhombic frame as a function of applied field, evaluated
for $|\Delta_{tet}/J|=0.6$ and $|\Delta_{o}/J|=0.4$. Panels a)-b)
show dominant contributions arising when the field is applied
along hard axis.} \label{fig:correlatorsHardCE}
\end{figure}
\subsubsection{Hard axis}
When the field is applied along the hard axis $b$, a stronger
intensity than that for the $a$ case is required to stabilize the
full PO region. The evolution towards PO is the result of the
smooth suppression of AFO correlations,as demonstrated by the
suppression of $\langle {\tau^a_y}_i {\tau^a_y}_j \rangle$ between
nearest neighbors [see Fig.~\ref{fig:correlatorsHardCE} a)]. In
the regime $0.4<|\Delta_{tet}/J|<0.6$, at the boundary between
AFO$^a_z$ and C-AFO$^a_y$, this determines a weakly
antiferro-orbital (WO) behavior, represented in
Fig.~\ref{fig:configCE} e). This region is the result of
disordering C-AFO$^a$ by introducing a nonvanishing onsite
momentum $\langle \tau^b_y \rangle$ along the hard axis direction.
As we already pointed out in previous section, this corresponds to
restore orbital fluctuations along $b$ by introducing $\hat{a}$
configurations, as illustrated in Fig.~\ref{fig:TauAction}.
Reactivating fluctuations in the whole orbital subspace, brings
the system back to
a situation where frustration of the orbital exchange is enhanced. \\
We notice that, contrary to the easy axis case, for the hard
axis orbital exchange between nearest neighbors is not
hindered. Moreover, we remark that the energy scale stabilizing WO
is now strongly related to the crystal field amplitudes. We explored the
evolution of region boundaries in Fig.~\ref{fig:PDHhardCE}, and
found that the onset of WO region grows linearly with $\Delta_o$.
These aspects underline that the softening of AFO correlations is
here caused by the reduction of onsite orbital momentum due to the new
activated flavor, instead of by the field-induced FO
correlations.

\section{Conclusions}

We have here presented a study of field tunability of orbital
correlations for layered ruthenates in an insulating ferromagnetic
configuration. We have analyzed the evolution of the possible
orbital patterns under the effect of the external magnetic field
coupled to the local angular momentum only. Numerical calculations
on a 2$\times$2 cluster and of a cluster embedded in a periodic
environment reveal that the competing effects driven by orbital
exchange and crystal field put the system on the verge of a FO/AFO
behavior, a circumstance that makes the character of OO in absence
of an applied field quite elusive. This outcome may be supported
by recent Resonant X-Ray scattering measurements on
Ca$_3$Ru$_2$O$_7$ ~\cite{Bohnenbuck 08}, where no appreciable
signal of staggered FO or AFO is found within experimental
resolution. We argue that this feature is at the origin of the
soft behavior and the easy field- tunability of the OO phases.
Particularly, AFO exchange leaves the place to strongnext-nearest
neighbors FO correlations when an external orbital field is
applied along the easy axis, while we get a tendency to a
isotropic weak antiferro-orbital correlated state for the hard
axis. An interpretation of these results can be summarized as
follows: the orthorhombic deformations tend to remove the orbital
degeneracy and lead to an ordered state where only two local
$t_{2g}$ configurations are active. The effect of the polarizing
field is to keep the orbital correlations in the two-orbital
sector when the field is along easy-axis; otherwise, along the
hard-axis the field forces the system to allow for all possible
local orbital fluctuations. The latter situation is responsible
for demolishing the OO ground state set by the competition between
Coulomb and crystal field interactions. Speaking on a general
ground, in many multi-orbital Mott insulators the metal-insulator
transition occurs as a symmetry breaking phase transition leading
to magnetic, orbital and charge orderings. We speculate that a
possible connection may exist between the demolition of long range
OO by an applied field and a likely metallic behavior and that
this connection may be of interest when referred to unconventional
magneto-conducting properties of Ca$_3$Ru$_2$O$_7$.\\ We finally
point out some important remarks about a quantitative connection
of our calculation to the physical parameters regime expected in
the layered ruthenates, and specifically of Ca$_3$Ru$_2$O$_7$.
According to our results, the evolution of region boundaries
between AFO, C-AFO or WO behaviors is mainly concentrated within
the windows $0.2 < \Delta_{tet}/J < 0.8$ and $0.1 < \Delta_{o}/J <
0.4$. Checking the range of the microscopic parameters that bias
the proposed orbital patterns is not easy since the amplitude of
$J$, $\Delta_{tet}$ and $\Delta_o$ is difficult to be directly
measured and require a model derivation. An explicit $ab-initio$
evaluation for Ca$_3$Ru$_2$O$_7$ is not yet available. For the
closest members of Ruddlesden-Popper series according to their
physical and structural properties, Ca$_2$RuO$_4$ and
Sr$_3$Ru$_2$O$_7$ respectively, one ends up with the following
$ab-initio$ estimates $J \sim 50 meV$, $\Delta_{tet} \sim 20 meV$
and $\Delta_o \sim 5 meV$~\cite{Gorelov10,Pavarini}. These values
set the ratios $\Delta_{tet}/J\sim 0.4$ and $\Delta_{o}/J\sim
0.1$, whose orders of magnitude are compatible with our choices of
analysis for the regions of phase diagrams where major changes
occur.


\begin{thebibliography}{99}
\bibitem{note} Here we refer to the notation adopted by Cao and
coworkers in ref~\cite{Cao00}. On the contrary, neutron
diffraction data~\cite{Yoshida05} and  X-ray scattering
studies~\cite{Nelson07} report $b$ as low temperature easy axis.
\end{thebibliography}
\end{document}